\documentclass[conference]{IEEEtran}
\usepackage[utf8]{inputenc}
\usepackage{bm}
\usepackage{amsmath,epsfig,amssymb,dsfont,amsthm}
\usepackage{hyperref}
\usepackage{verbatim}
\usepackage{caption}
\usepackage{subcaption}
\usepackage{xcolor}
\usepackage{graphicx}
	\graphicspath{{./figures/}}
\usepackage{float}
\usepackage{url}
\usepackage{eurosym}
\usepackage{setspace}
\usepackage{balance}
\usepackage{dsfont}
\usepackage{bbm}
\usepackage{array}
\usepackage{cite}
\usepackage[ruled,vlined]{algorithm2e}

\newcommand{\blambda}{\boldsymbol{\Lambda}}
\newcommand{\bA}{\boldsymbol{A}}
\newcommand{\bL}{\boldsymbol{\mathcal L}}
\newcommand{\R}{\mathcal R}
\newcommand{\bT}{\mathsf{T}}
\newcommand{\bE}{\mathbb{E}}

\theoremstyle{plain}
\newtheorem{Thm}{Theorem}

\newtheorem{Lem}{Lemma}
\newtheorem{Asm}{Assumption}
\theoremstyle{remark}

\pdfoutput=1

\newcommand{\qedsymb}{\hfill\ensuremath{\blacksquare}}                 
\allowdisplaybreaks
\title{Online Graph Learning from Social Interactions}

\author{\IEEEauthorblockN{Valentina Shumovskaia\IEEEauthorrefmark{1}, Konstantinos Ntemos\IEEEauthorrefmark{1}, Stefan Vlaski\IEEEauthorrefmark{2} and Ali H. Sayed\IEEEauthorrefmark{1}}
\IEEEauthorblockA{\IEEEauthorrefmark{1}
École Polytechnique Fédérale de Lausanne (EPFL)
}
\IEEEauthorblockA{\IEEEauthorrefmark{2}
Imperial College London}
\thanks{Emails: $\{$valentina.shumovskaia, konstantinos.ntemos, ali.sayed$\}$@epfl.ch, s.vlaski@imperial.ac.uk. This work was supported in part by SNSF grant 205121-184999.}
}

\IEEEoverridecommandlockouts
\begin{document}

\maketitle

\begin{abstract}
    Social learning algorithms provide models for the formation of opinions over social networks resulting from local reasoning and peer-to-peer exchanges. Interactions occur over an underlying graph topology, which describes the flow of information and relative influence between pairs of agents. For a given graph topology, these algorithms allow for the prediction of formed opinions. In this work, we study the inverse problem. Given a social learning model and observations of the evolution of beliefs over time, we aim at identifying the underlying graph topology. The learned graph allows for the inference of pairwise influence between agents, the overall influence agents have over the behavior of the network, as well as the flow of information through the social network. The proposed algorithm is online in nature and can adapt dynamically to changes in the graph topology or the true hypothesis.
\end{abstract}


\section{Introduction}
    Graphs provide a useful tool to model and exploit relations in high-dimensional data, such as social networks~\cite{barnes1969graph, newman2002random}, roadway networks~\cite{cui2019traffic, zheng2020gman}, and molecular data~\cite{chen2019graph, kearnes2016molecular}, among others. A common observation in all these settings is the fact that the graph topology impacts data distribution and evolution. Hence, knowledge of the graph has the potential to improve the performance of inference tasks. However, the underlying graph structure is unknown in many applications and needs to be estimated through observations. A number of solutions for graph learning~\cite{kalofolias2016learn, egilmez2017graph, vlaski2018online, dong2019learning, pasdeloup2017characterization, thanou2017learning, chepuri2017learning, shafipour2017network, segarra2016network, viola2018graph, sardellitti2016graph, maretic2017graph} have already been proposed in the literature, where algorithms for graph inference have been developed for particular models, describing the relationship between observations and graphs.
    
    For instance, graph learning for the heat diffusion process is studied in~\cite{pasdeloup2017characterization, vlaski2018online, thanou2017learning, ma2008mining}, while learning under structural constraints, such as connectivity~\cite{egilmez2017graph} and sparsity, appears in~\cite{egilmez2017graph, chepuri2017learning, maretic2017graph}, and approaches based on examining the precision matrix appear in  \cite{friedman2008sparse, matta2019graph}. Most of these works consider static graphical models. This is in contrast to graphs with dynamic properties~\cite{vlaski2018online} where the connectivity among agents can change over time.
    
    In this work, we develop an algorithm for graph learning in the social learning setting {\color{black}where agents react to streaming data and also to information shared with their neighbors}.
    Our study focuses on the social learning paradigm studied in earlier works \cite{jadbabaie2012non, nedic2017fast, molavi2017foundations, molavi2018theory, bordignon2020adaptive, bordignon2020social, lalitha2018social, zhao2012learning, gale2003bayesian, acemoglu2011bayesian}. The combination weights are unknown and need to be inferred through observations acquired during the agents' learning process.
    
    Social learning refers to the problem of distributed hypothesis testing, where each agent aims at learning an underlying true hypothesis (or state) {\color{black}through its own observations and from information shared by its neighbors}. Social learning studies can be categorized into Bayesian \cite{gale2003bayesian, acemoglu2011bayesian} and non-Bayesian \cite{jadbabaie2012non, nedic2017fast, molavi2017foundations, molavi2018theory, bordignon2020adaptive, bordignon2020social, lalitha2018social, zhao2012learning}. Non-Bayesian approaches have gained increased interest due to their appealing scalability traits. In these approaches, at every time instant, agents follow a two-stage process. First, every agent updates its belief (which is a probability distribution over the possible hypotheses) based on its current received observation. Then, it fuses the shared beliefs from its neighbors. The main focus of these studies is to prove that agents' beliefs across the network converge to the true hypothesis after sufficient repeated interactions.
    
    In this work, we are interested in revealing the underlying influence pattern. More specifically, we formulate the inverse problem, where given the evolution of beliefs, the objective is to identify the relative influence between pairs of agents, captured by the graph topology. We are interested in studying a dynamic setting where both the graph topology and the true hypothesis can change over time.  Therefore, we consider that the agents follow the adaptive social learning protocol \cite{bordignon2020adaptive}.
    
    We describe the system model in Section~\ref{sec:system_model}, while Section~\ref{sec:algorithm} describes the algorithm and examines 
    the steady-state performance of the graph learning process.
    In Section~\ref{sec:experiments}, we provide experiments and illustrate the robustness of the proposed method against dynamic changes on the graph topology.

\section{Social Learning Model}
\label{sec:system_model}
    We consider a set $\mathcal{N}$ of agents connected by a graph $\mathcal{G}=\langle \mathcal{N}, \mathcal{E} \rangle$, where $\mathcal{E}$ represents the links between agents. Two agents that are linked can exchange information directly with each other. The set of neighbors of an agent $k\in\mathcal{N}$ including itself, is denoted by $\mathcal{N}_k$.
    
    All agents aim at learning the true hypothesis $\theta^{\star}$, belonging to a set of all possible hypotheses denoted by $\Theta$ (whose cardinality is at least two). 
    To this end, each agent $k$ has access to observations $\boldsymbol{\zeta}_{k,i}\in\mathcal{Z}_k$ at every time $i\geq1$. Agent $k$ also has access to the likelihood functions $L_k(\boldsymbol{\zeta}_{k,i}|\theta)$, for all $\theta\in\Theta$. 
    The signals $\boldsymbol{\zeta}_{k,i}$ are independent over both time and space, and are also identically distributed (i.i.d.) over time. 
    We will use the notation $L_k(\theta)$ instead of $L_k(\boldsymbol{\zeta}_{k,i}\vert\theta)$ for brevity.
    At each time $i$, agent $k$ keeps a {\em belief vector} $\boldsymbol{\mu}_{k,i}$, which is a probability distribution over the possible states. The belief component $\boldsymbol{\mu}_{k,i}(\theta)$ quantifies the confidence of agent $k$ that $\theta$ is the true state. Therefore, at time $i$, each agent's true state estimator is as follows:
    \begin{align}
        \widehat{\boldsymbol{\theta}}_{k,i}^\circ = \arg\max_{\theta\in\Theta}\boldsymbol{\mu}_{k,i}(\theta).
        \label{eq:truestate_est0}
    \end{align}  
    To avoid technicalities, where agents discard a particular state a priori, we impose the following assumption on initial beliefs.
    \begin{Asm} {\bf (Positive initial beliefs).}
        \label{positive_beliefs}
            For all hypotheses $\theta\in\Theta$, all agents $k\in\mathcal N$ start with positive initial belief $\boldsymbol{\mu}_{k,0}(\theta)>0$.
        \qedsymb
        \label{asm:beliefs}
    \end{Asm}
    At every time instant $i$, every agent $k$ updates its belief by using a two-stage process. First, it incorporates information from the received observation $\boldsymbol{\zeta}_{k,i}$ and then it fuses the information from its neighbors. More specificially, in this work we consider the \textit{adaptive social learning} rule~\cite{bordignon2020adaptive}, which has been shown to have favorable transient and steady-state performance in terms of convergence rate and probability of error. Under this protocol, agents update their beliefs in the following manner:
    \begin{align}
        &\boldsymbol{\psi}_{k,i}(\theta)=
        \frac{L_k^\delta(\boldsymbol{\zeta}_{k,i}|\theta)\boldsymbol{\mu}^{1-\delta}_{k,i-1}(\theta)}{\sum_{\theta'\in\Theta}L_k^{\delta}(\boldsymbol{\zeta}_{k,i}|\theta')\boldsymbol{\mu}^{1-\delta}_{k,i-1}(\theta')},\quad k\in\mathcal{N}\label{eq:adapt_adaptive}\\
        \nonumber\\
        &\boldsymbol{\mu}_{k,i}(\theta)=\frac{\prod_{\ell\in\mathcal{N}_k}\boldsymbol{\psi}^{a_{\ell k}}_{\ell,i}(\theta)}{\sum_{\theta'\in\Theta}\prod_{\ell\in\mathcal{N}_k}\boldsymbol{\psi}^{a_{\ell k}}_{\ell,i}(\theta')}, \quad k\in\mathcal{N} \label{eq:combine}
    \end{align}
    where $a_{\ell k}$ denotes the {\em combination weight} assigned by agent $k$ to neighboring agent $\ell$, satisfying $0<a_{\ell k}\leq1$, for all $\ell\in\mathcal{N}_k$, $a_{\ell k}=0$ for all $\ell\notin\mathcal{N}_k$, and $\sum_{\ell\in\mathcal{N}_k}a_{\ell k}=1$. The algorithm is called ``adaptive'' due to the step-size parameter $\delta \in (0,1)$, which allows it to track changes in the true hypothesis $\theta^\star$. Observe that the numerator in~(\ref{eq:combine}) is the weighted geometric mean of the priors $\bm{\psi}_{\ell,i}(\theta)$ at time $i$ with weights given by the scalars $\{a_{\ell k}\}$.
    
    Let $A_{\star}$ denote the left-stochastic {\em combination matrix} consisting of all combination weights $a_{\ell k}$. 
    Regarding the network topology, we impose the following assumption~\cite{lalitha2016social, nedic2017fast, bordignon2020adaptive}, which allows information to flow throughout the whole network.
    \begin{Asm}{\bf{(Strongly-connected network)}.}
        \label{strongly_connected}
            The communication graph is {\em strongly connected} (i.e., there exists a path with positive weights linking any two agents, and at least one agent in the graph has a self-loop, meaning that there is at least one agent $k\in\mathcal{N}$ with $a_{kk}>0$).
        \qedsymb
        \label{asm:network}
    \end{Asm}
    
    Finally, we impose assumptions on the agents' observation models. We assume that the agents can collectively identify the underlying true hypothesis~\cite{lalitha2016social, bordignon2020adaptive}.
    \begin{Asm}{\bf{(Identifiability assumption)}.}
            For each wrong hypothesis $\theta\neq\theta^\star$, there is at least one agent $k\in \mathcal N$ that has strictly positive KL-divergence $D_{KL}\left(L_k\left(\theta\right)||L_k\left(\theta_\star\right)\right) > 0$.
        \newline $\textrm{ }$\qedsymb
        \label{asm:ident}
    \end{Asm}
    We also assume the boundedness of the likelihood functions~\cite{bordignon2020social}.
    \begin{Asm}{\bf{(Bounded likelihoods).}}
        \label{asm:support}
        There is a finite constant $b > 0$ such that, for all $k \in \mathcal N$:
        \begin{align}
            \Bigg|\log \frac {L_k(\boldsymbol\zeta | \theta)}{L_k(\boldsymbol\zeta | \theta')} \Bigg| \leq b
        \end{align}
        for all $\theta,\;\theta' \in \Theta$, and $\boldsymbol\zeta \in \mathcal Z_k$.
        \qedsymb
    \end{Asm}
    
\section{Inverse Modeling Problem}
\label{sec:algorithm}
    \subsection{Problem Statement}
        In our study, we assume that the graph is completely hidden. The assumption is motivated by the fact that in real-world settings, the pattern of interactions among agents is usually unknown to an external observer. In addition, in the social learning strategy, it is common~\cite{lalitha2016social, lalitha2018social} to assume that for each time $i\geq 1$, each agent local observation $\boldsymbol{\zeta}_{k,i}$ is private and external observers do not have access to it. On the other hand, beliefs (i.e., $\boldsymbol{\psi}_{k,i}(\theta)$) are public and exchanged across the network. For this reason, our goal is to infer the graph topology by observing the exchanged beliefs among the agents.
        
        Formally, we assume that at each time step $i\geq 1$ we observe the beliefs of the agents in the network, collected into the set:
        \begin{align}
        \label{obs_information}
            {\mathcal D}_i = \Bigl\{ \bm{\psi}_{k,i}(\theta),\;k\in{\mathcal N}\Bigr\}
        \end{align}
        The problem of interest is to recover the combination matrix $A_\star$ based on knowledge of $\{\mathcal D_i\}_{i\geq 1}$.
        
    \subsection{Likelihood and Beliefs Ratios}
        We define the matrices $\blambda_i$ and $\bL_i$ of size $|\mathcal N| \times (|\Theta|-1)$, where each element is a relative measure of log beliefs and likelihood ratios as follows:
        \begin{align}
            &[\boldsymbol{\Lambda}_{i}]_{k,j} \triangleq\log\frac{\boldsymbol{\psi}_{k,i}(\theta_0)}{\boldsymbol{\psi}_{k,i}(\theta_j)}
            \label{eq:lambda}\\
            &[\boldsymbol{\mathcal{L}}_{i}]_{k,j}\triangleq\log\frac{L_k(\boldsymbol{\zeta}_{k,i}\vert\theta_0)}{L_k(\boldsymbol{\zeta}_{k,i}\vert\theta_j)},
            \label{eq:loglikelihood}
        \end{align}
        In these expressions, we have chosen some $\theta_0\in\Theta$ as a reference state, while $\theta_j\neq \theta_0$. Due to Assumption~\ref{asm:support}, $\bL_i$ has finite entries. 
        Since we initially start from positive beliefs $\boldsymbol\mu_{k,0}(\theta)$, and the likelihoods remain positive with probability one, it follows from the update rules (\ref{eq:adapt_adaptive})-(\ref{eq:combine}) that $\blambda_i < \infty$.
        
        Observe that both matrices vary with  the time index $i$. Based on the definitions~(\ref{eq:lambda})-(\ref{eq:loglikelihood}), some algebra will show that we can transform~(\ref{eq:adapt_adaptive})-(\ref{eq:combine}) into an update relating these matrices:
        \begin{align}
            &\boldsymbol{\Lambda}_i=(1-\delta)A_\star^{\mathsf{T}}\boldsymbol{\Lambda}_{i-1}+\delta\boldsymbol{\mathcal{L}}_i.
            \label{eq:recursion_adaptive}
        \end{align}
        
        At every iteration $i$, the quantities $\{\blambda_i,\blambda_{i-1}\}$ are known based on knowledge of the beliefs ${\mathcal D}_i$ from (5). On the other hand, the quantity $\bL_i$ is not known because the observations $\{\bm{\zeta}_{k,i}\}$ are private.  We wish to devise a scheme that allows us to estimate $A_{\star}$ in (8) from knowledge of $\{\blambda_i,\blambda_{i-1}\}$ and from a suitable approximation for $\bL_i$. Before discussing the learning algorithm, however, we establish the following useful property. For simplicity of notation, we will write 
        \begin{align}
            \mathbb E [\cdot] \triangleq \mathbb E_{\boldsymbol\zeta_{k,t}\sim L_k(\theta^\star), k\in\mathcal N, t\leq i} [\cdot]
        \end{align}
        where the expectation is relative to the randomness in all local observations up to time $i$. 
        
        \begin{Lem}[\bf{Mean likelihood matrix}]
            Random variables $\bL_i$ are i.i.d. over time and space, and their mean matrix $\bar\bL=\bE\bL_i$ is independent of time and finite with each entry equal to:
            \begin{align}
                [\bar \bL]_{k,j} =\textrm{ }&  D_{KL}\left(L_k\left(\theta^\star\right)||L_k\left(\theta_j\right)\right) \nonumber\\
                &- D_{KL}\left(L_k\left(\theta^\star\right)||L_k\left(\theta_0\right)\right).
                \label{eq:L_exp}
            \end{align}
            \label{lem:loglikelihoods}
        \end{Lem}
        \begin{proof}
            Proof is omitted due to space limitations.
        \end{proof}
        
    \subsection{Algorithm Development}

        The linear nature of the update for $\blambda_i$ in~(\ref{eq:recursion_adaptive}) motivates the following instantaneous quadratic loss function for finding $A_\star$:
        \begin{align}
            Q'(A; \blambda_i, \blambda_{i-1}) = \frac{1}{2} \| \boldsymbol{\Lambda}_i -(1-\delta)A^{\mathsf{T}}\boldsymbol{\Lambda}_{i-1} - \delta \boldsymbol{\mathcal{L}}_i \|_{\rm F}^2,
            \label{eq:cost0}
        \end{align}
        where $\|\cdot\|_{\rm F}$ denotes Frobenius norm. Computation of $\bL_i$ requires knowledge of $\boldsymbol\zeta_{k, i}$, $k\in\mathcal N$, which is assumed to be private for each agent, therefore hidden from the observer. For this reason, we will assume only knowledge of $\bar \bL$, which is in principle requires knowledge of the true hypothesis $\theta^\star$ due to Lemma~\ref{lem:loglikelihoods}. We explain in the sequel how to circumvent this requirement. 
        
        Typically, at each time step $i\geq 1$, every agent $k\in\mathcal N$ estimates the true state using~(\ref{eq:truestate_est0}).
        It can be shown \cite[Theorem 2]{bordignon2020adaptive}) that the probability of error $\mathbb P (\widehat{\boldsymbol{\theta}}_{k,i}^\circ \neq \theta^\star) \rightarrow 0$ as $i\rightarrow\infty$ and $\delta \rightarrow 0$. It can be verified that the same conclusion continues to hold if we estimate the underlying hypothesis based on the intermediate belief vectors (which are the quantities that are assumed to be observable):
        \begin{align}
            \widehat{\boldsymbol{\theta}}_{k,i} = \arg\max_{\theta\in\Theta}\boldsymbol{\psi}_{k,i}(\theta).
        \end{align}

        In order to have agreement on the $\theta_\star$ among the agents, we will estimate a common $\widehat{\boldsymbol{\theta}}_{i} $ by using a majority vote rule. Then, the following conclusion holds.
        \begin{Lem}[\bf{True state learning error: majority vote}]
            \begin{align}
                \lim_{\delta\rightarrow0}\mathbb P \left(\lim_{i\rightarrow\infty}\widehat {\boldsymbol\theta}_{i} \neq \theta^\star\right)  = 0.
            \end{align}
            \label{lem:majvote}
        \end{Lem}
        \begin{proof}
            Proof is omitted due to space limitations.
        \end{proof}
        
        Therefore, we replace~(\ref{eq:cost0}) by the following loss function:
        \begin{align}\label{eq:cost_function_adaptive} 
            &Q(A; \blambda_i, \blambda_{i-1}, \bar \bL_i) = \frac{1}{2} \| \boldsymbol{\Lambda}_i -(1-\delta)A^{\mathsf{T}}\boldsymbol{\Lambda}_{i-1} - \delta \bar \bL_i \|_{\rm F}^2.
        \end{align}
        with $\bar\bL_i = \bE_{\widehat{\boldsymbol\theta}_i}\bL_i$, and $\mathbb E_{\widehat{\boldsymbol\theta}_i}$ means that the expectation is computed assuming that the private data $\boldsymbol{\zeta}_{k,i}$ is generated according to $\widehat{\boldsymbol\theta}_i:$ $\boldsymbol{\zeta}_{k,i} \sim L_k(\boldsymbol{\zeta}_{k,i}|\widehat{\boldsymbol\theta}_i)$.
        Our minimization problem over a horizon of $N$ observations then becomes:
        \begin{align}
        \label{eq:min_problem}
            &\min_{A} J(A) \triangleq \frac 1N\sum_{i=1}^N J_i(A),\\
            &J_i(A) \triangleq \bE Q(A;\blambda_{i}, \blambda_{i-1}, \bar\bL_i)
        \end{align}
        where the statistical properties of $\blambda_i$ vary with time. This explains why we are averaging over a time-horizon in~(\ref{eq:min_problem}).
        We apply stochastic approximation to solve~(\ref{eq:min_problem}), namely, a recursion of the form:
        \begin{align}
            \boldsymbol{A}^{\mathsf{T}}_i =&\textrm{ } \boldsymbol{A}^{\mathsf{T}}_{i-1}
            + \mu(1-\delta) \nonumber\\
            \;\times&\left(\boldsymbol{\Lambda}_i -  (1-\delta)\boldsymbol{A}_{i-1}^{\mathsf{T}}\boldsymbol{\Lambda}_{i-1} - \delta \bar \bL_i \right) \boldsymbol{\Lambda}_{i-1}^{\mathsf{T}}
            \label{eq:graph_update_unbiased}
        \end{align}

        In order to examine the steady-state performance of the algorithm, we introduce an independence assumption that is common in the study of adaptive systems~\cite{Sayed_2014}.
        \begin{Asm}
            {\bf{(Separation principle)}}
            \label{asm:independence}
            Let $\widetilde{\bA}_i=A_\star-\bA_i$ denote the estimation error. Assume the step-size $\mu$ is sufficiently small, so that in the limit, $\|\widetilde{\bA}_{i}\|_{\rm F}^2$ reaches a steady state distribution, and $\widetilde{\bA}_i$ is independent of $\blambda_i$.
            \qedsymb
        \end{Asm}

        \noindent Using this condition, we can establish the following steady-state performance for the Online Graph Learning (OGL) algorithm.
        \begin{Thm}[\bf{Steady-state performance}]
            Under Assumptions
            \ref{asm:beliefs}-\ref{asm:independence}, after large enough number of social learning iterations with $\delta \rightarrow 0$ and for sufficiently small $\mu$, the mean squared deviation converges exponentially fast with:
            \begin{align}
                \limsup_{i\rightarrow\infty} \bE \|\widetilde{\bA}_{i}\|_{\rm F}^2
                \leq \frac{\mu^2\gamma}{1-\alpha} = O(\mu),
            \end{align}
            where
            \begin{align}
                \alpha &= 1 - 2\mu\nu + O(\mu^2) \nonumber\\
                \gamma &=\delta^2\kappa |\mathcal N|\lambda_{\max}(\R_{\bL})\nonumber\\
                \nu &= (1-\delta)^2\lambda_{\min}\left(\R_{\blambda}\right)\nonumber\\
                \kappa &= (1-\delta)^2\lambda_{\max}\left(\R_{\blambda}\right)
            \end{align}
            and $\R_{\bL} \triangleq \bE (\bL_i - \bar \bL)(\bL_i - \bar\bL)^\bT$ is independent of $i$, whereas $\R_{\blambda} \triangleq \lim_{i\rightarrow\infty} \bE\blambda_i\blambda_i^\bT$ is finite.
            \label{thm:conv_0}
        \end{Thm}
        \begin{proof}
            Proof is omitted due to space limitations.
        \end{proof}

\section{Computer Simulations}
\label{sec:experiments}

    The experiments that follow help illustrate the ability of the proposed algorithm to identify edges and to adapt to situations where the graph topology is dynamic, as well as the hypothesis.
    
    \subsection{Setup}
        We consider a network of $30$ agents with $|\Theta|=4$ states, where the adjacency matrix is generated according to the Erdos-Renyi model with edge probability $p=0.2$. We set $\mathcal Z_k$ to be a discrete sample space with $|\mathcal Z_k| = 4$ for $k\in\mathcal N$. The step-size of the model is set to $\delta=0.05$. We define the likelihood functions $L_k(\theta)$, $k\in\mathcal N$, $\theta\in\Theta$ as follows:
        \begin{align}
            &L_k(\bm\zeta|\theta) = \sum_{z \in \mathcal Z_k}\mathbb I[\bm\zeta = z]\beta_{k,z}(\theta),\nonumber\\
            &\beta_{k,z}(\theta) \geq 0, z \in \mathcal Z_k\nonumber\\
            &\sum_{z \in \mathcal Z_k}\beta_{k,z}(\theta) = 1,
        \end{align}
        where the parameters $\beta_{k,z}(\theta)$ are generated randomly. During the graph learning procedure, we use $\mu=0.01$.
    
    \subsection{Graph learning}
\begin{figure}
            \centering
            \begin{subfigure}[b]{0.475\textwidth}
                \centering
                \includegraphics[width=0.9\textwidth]{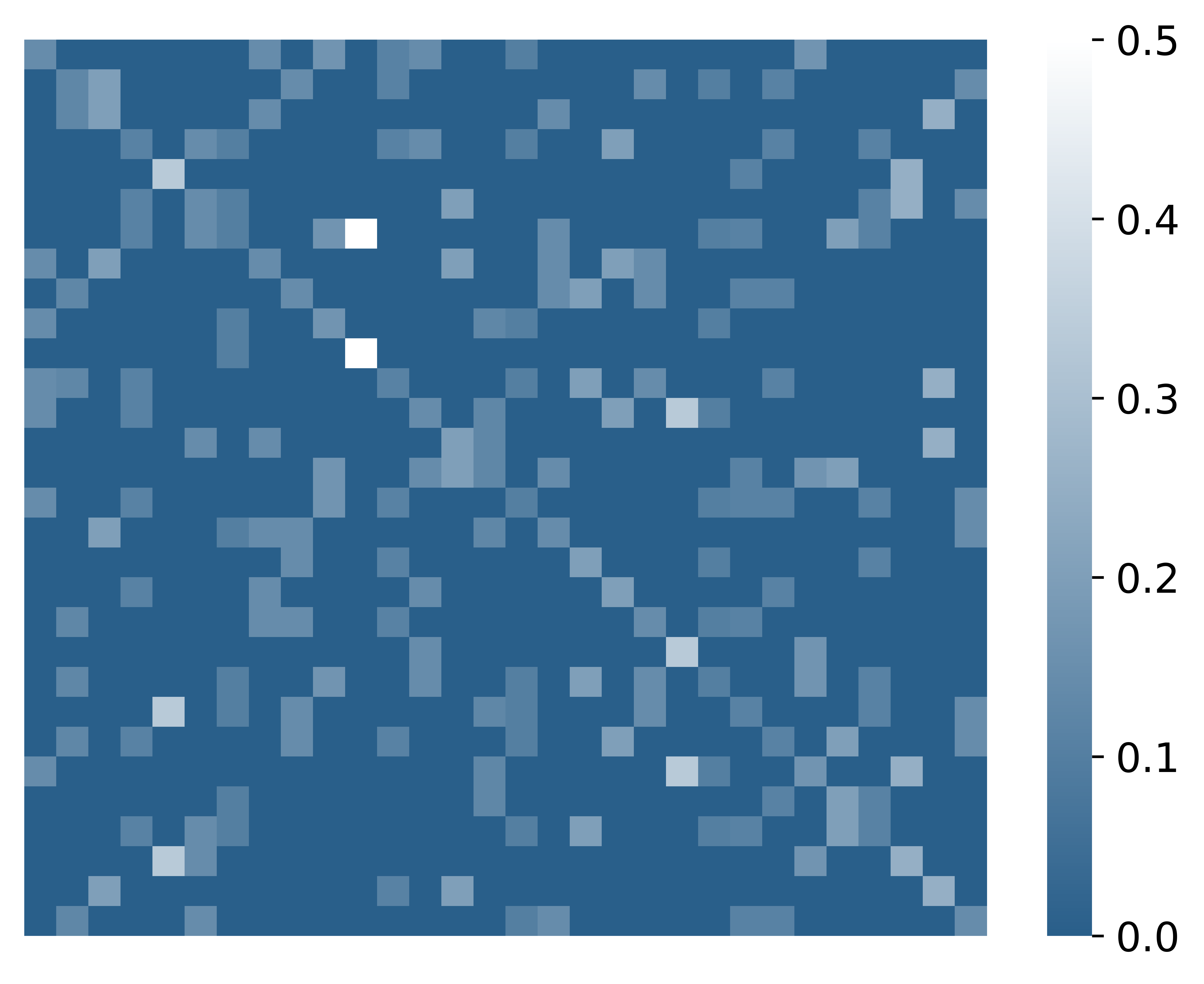}
                \subcaption{True graph.}
                \label{fig:adj_true}
            \end{subfigure}
            \vfill
            \begin{subfigure}[b]{0.475\textwidth}  
                \centering 
                \includegraphics[width=0.9\textwidth]{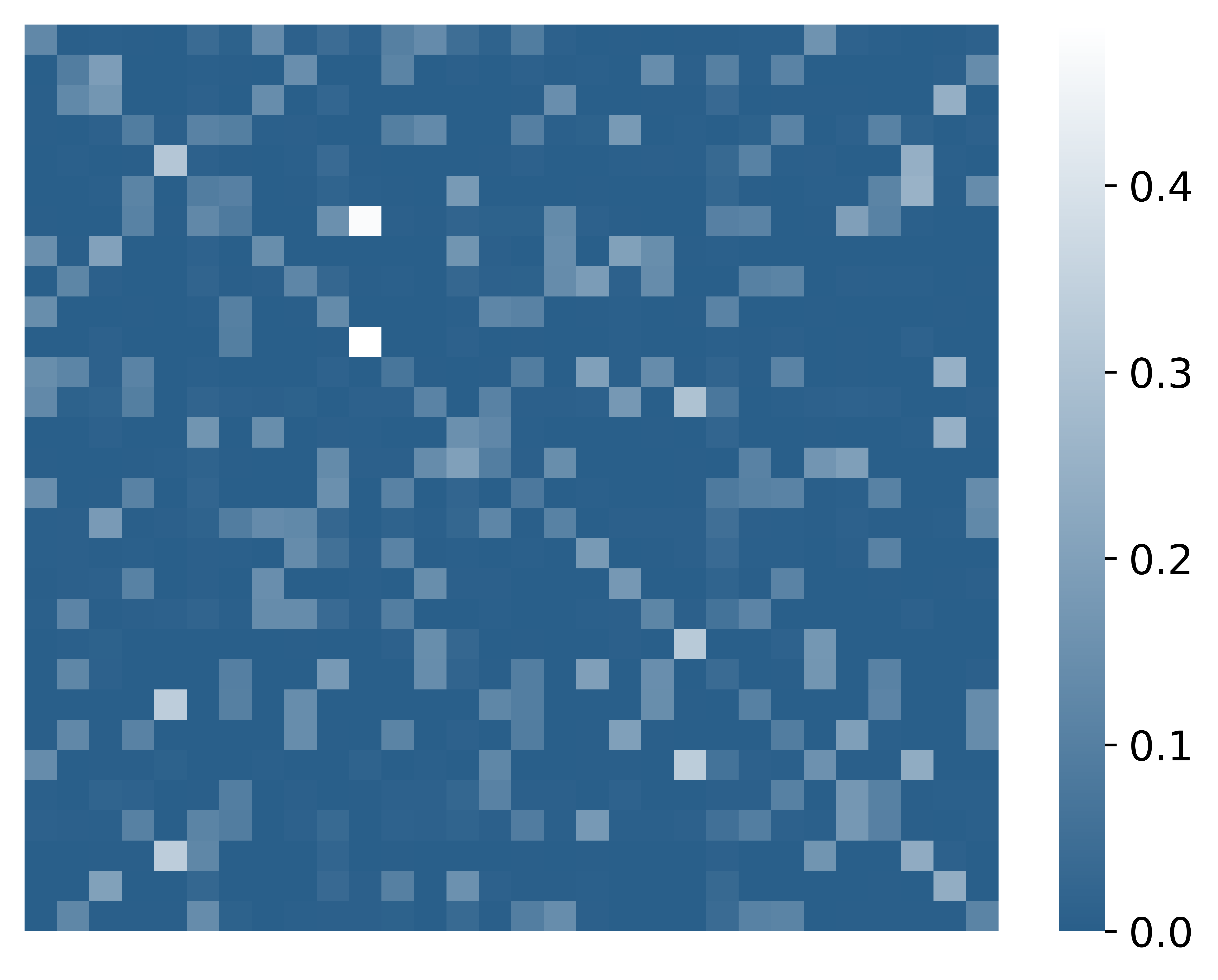}
                \subcaption{Learned graph.}
                \label{fig:adj_learned}
            \end{subfigure}
            \caption{True combination matrix and the learned matrix using the Online Graph Learning (OGL) algorithm.}
            \label{fig:adj}
        \end{figure}

        We provide a comparison between the true combination matrix and the estimated combination matrix. We plot the combination matrices in Fig.~\ref{fig:adj}. 
        The experiment shows the ability of the algorithm to identify the graph: the recovered combination weights are close to the actual weights. In no-edge places, we observe reasonably small weights on the recovered matrix. These can be removed in post-processing by simple thresholding or more elaborate schemes, such as the $k$-means algorithm~\cite{matta2020graphlearning}.
        
\begin{figure}
            \centering
            \includegraphics[width=0.475\textwidth]{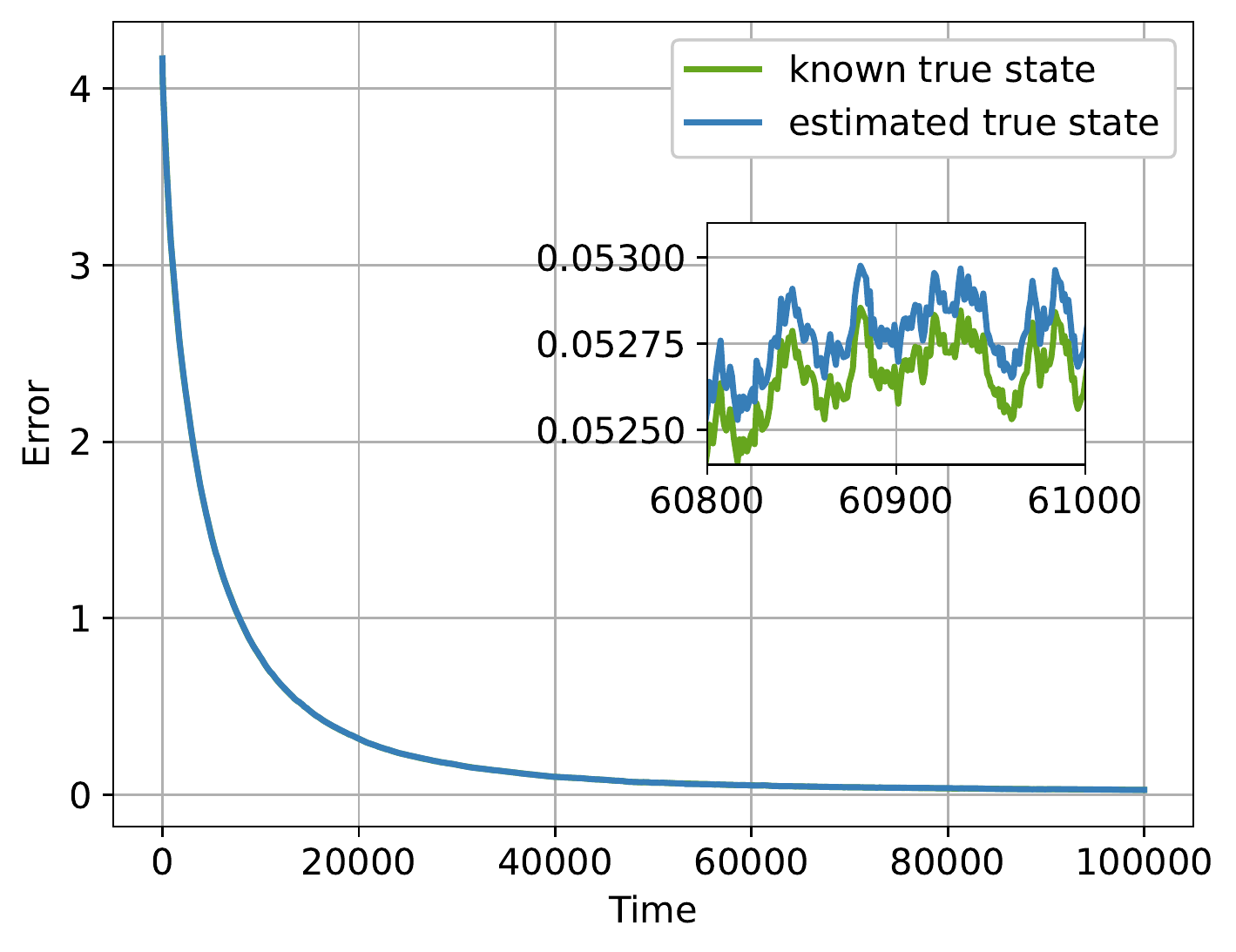}
            \caption{Error evolution of two algorithm variants: known true state and estimated true state.}
            \label{fig:error_rate}
        \end{figure}

        Additionally, in Fig.~\ref{fig:error_rate}, we plot how the deviation from the true matrix $A_\star$ evolves. The deviation is computed as the following quantity: 
        \begin{align}
            \|\widetilde{\bA}_i\|_{\rm F}^2 = \|A_\star - \bA_i\|_{\rm F}^2.
        \end{align} 
        We provide the error rates for both algorithm variants with known true state $\theta^\star$ and estimated true state $\widehat{\boldsymbol\theta}_i$. We see that there is a negligible gap between the learning performances.
        
\begin{figure}
            \centering 
            \includegraphics[width=0.475\textwidth]{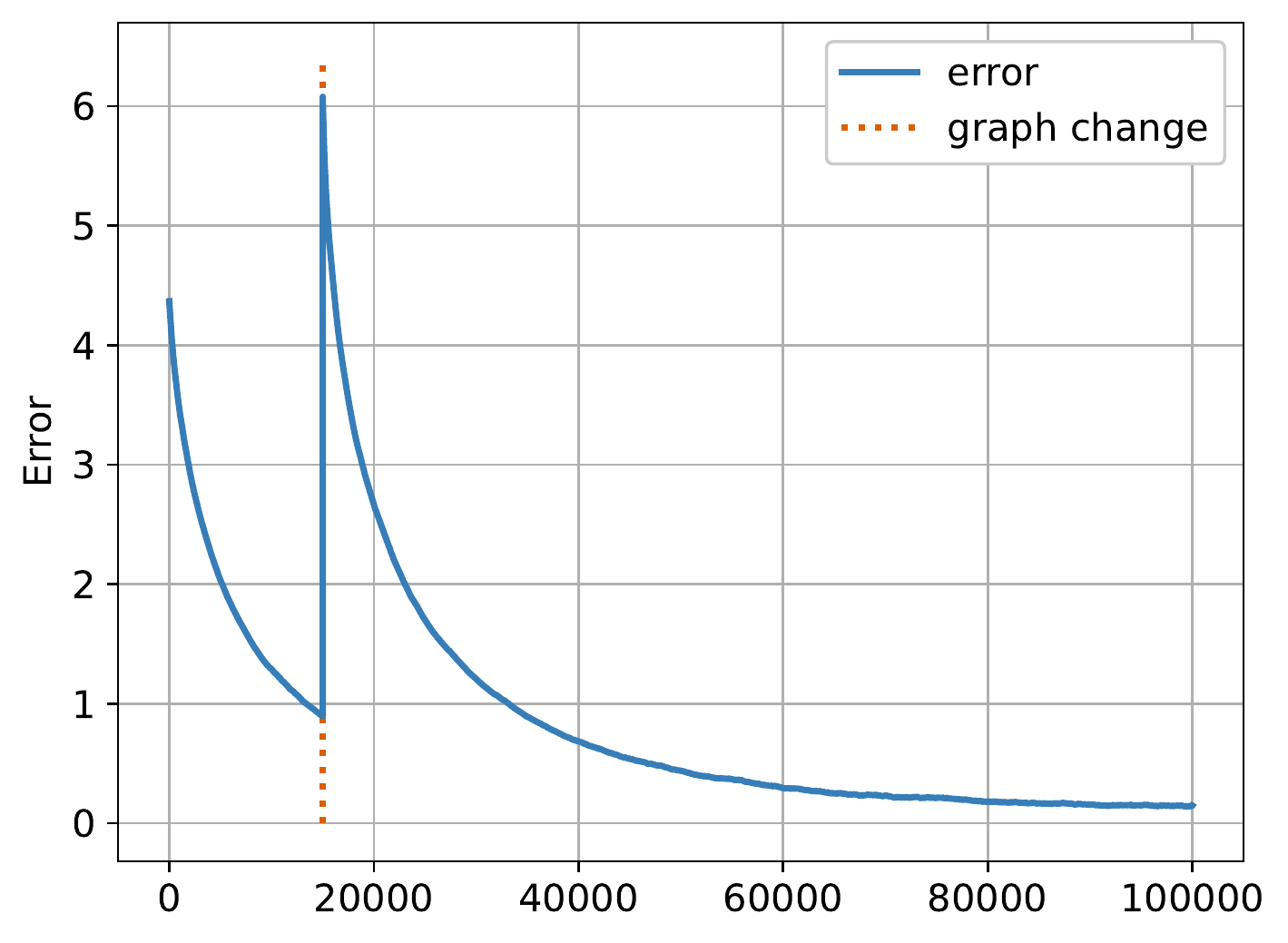}
            \caption{Error evolution when the graph edges are regenerated.}
            \label{fig:graph_change}
        \end{figure}

        The proposed algorithm is robust to changes in the true state and graph topology.
        In Fig.~\ref{fig:graph_change}, we regenerate edges at time $15000$. The algorithm adapts and converges to the new combination matrix at a linear rate. 
        Thus, we have experimentally illustrated that the algorithm is stable to dynamic network changes, which is a natural setting to consider in practice. These properties hold because the algorithm is online and processes data one by one with a constant learning rate $\mu>0$.

\section{Conclusions}
    In this paper, the problem of graph learning through observing social interactions is investigated. We develop an online algorithm that learns the agents' influence pattern via observing agents' beliefs over time. We prove that the proposed algorithm successfully learns the underlying combination weights matrix and demonstrate its performance through analysis and computer simulations. In this way, we are able to discover the pattern of information flow in the network. A distinct feature of the proposed algorithm is the fact that it can track changes in the graph topology as well as in the true hypothesis. 
    
    As future work, we aim to investigate {\em partial information} setting, where the algorithm has access only to the beliefs of a subset of the network agents.
\bibliographystyle{IEEEtran}
\bibliography{main}

\begin{thebibliography}{10}
\providecommand{\url}[1]{#1}
\csname url@samestyle\endcsname
\providecommand{\newblock}{\relax}
\providecommand{\bibinfo}[2]{#2}
\providecommand{\BIBentrySTDinterwordspacing}{\spaceskip=0pt\relax}
\providecommand{\BIBentryALTinterwordstretchfactor}{4}
\providecommand{\BIBentryALTinterwordspacing}{\spaceskip=\fontdimen2\font plus
\BIBentryALTinterwordstretchfactor\fontdimen3\font minus
  \fontdimen4\font\relax}
\providecommand{\BIBforeignlanguage}[2]{{%
\expandafter\ifx\csname l@#1\endcsname\relax
\typeout{** WARNING: IEEEtran.bst: No hyphenation pattern has been}%
\typeout{** loaded for the language `#1'. Using the pattern for}%
\typeout{** the default language instead.}%
\else
\language=\csname l@#1\endcsname
\fi
#2}}
\providecommand{\BIBdecl}{\relax}
\BIBdecl

\bibitem{barnes1969graph}
J.~A. Barnes, ``Graph theory and social networks: A technical comment on
  connectedness and connectivity,'' \emph{Sociology}, vol.~3, no.~2, pp.
  215--232, 1969.

\bibitem{newman2002random}
M.~E. Newman, D.~J. Watts, and S.~H. Strogatz, ``Random graph models of social
  networks,'' \emph{Proc. National Academy of Sciences}, vol.~99, no. suppl 1,
  pp. 2566--2572, 2002.

\bibitem{cui2019traffic}
Z.~Cui, K.~Henrickson, R.~Ke, and Y.~Wang, ``Traffic graph convolutional
  recurrent neural network: A deep learning framework for network-scale traffic
  learning and forecasting,'' \emph{IEEE Transactions on Intelligent
  Transportation Systems}, vol.~21, no.~11, pp. 4883--4894, 2019.

\bibitem{zheng2020gman}
C.~Zheng, X.~Fan, C.~Wang, and J.~Qi, ``Gman: A graph multi-attention network
  for traffic prediction,'' in \emph{Proceedings of the AAAI Conference on
  Artificial Intelligence}, vol.~34, no.~01, 2020, pp. 1234--1241.

\bibitem{chen2019graph}
C.~Chen, W.~Ye, Y.~Zuo, C.~Zheng, and S.~P. Ong, ``Graph networks as a
  universal machine learning framework for molecules and crystals,''
  \emph{Chemistry of Materials}, vol.~31, no.~9, pp. 3564--3572, 2019.

\bibitem{kearnes2016molecular}
S.~Kearnes, K.~McCloskey, M.~Berndl, V.~Pande, and P.~Riley, ``Molecular graph
  convolutions: moving beyond fingerprints,'' \emph{Journal of Computer-Aided
  Molecular Design}, vol.~30, no.~8, pp. 595--608, 2016.

\bibitem{kalofolias2016learn}
V.~Kalofolias, ``How to learn a graph from smooth signals,'' in
  \emph{Artificial Intelligence and Statistics}.\hskip 1em plus 0.5em minus
  0.4em\relax PMLR, 2016, pp. 920--929.

\bibitem{egilmez2017graph}
H.~E. Egilmez, E.~Pavez, and A.~Ortega, ``Graph learning from data under
  laplacian and structural constraints,'' \emph{IEEE Journal of Selected Topics
  in Signal Processing}, vol.~11, no.~6, pp. 825--841, 2017.

\bibitem{vlaski2018online}
S.~Vlaski, H.~P. Maretić, R.~Nassif, P.~Frossard, and A.~H. Sayed, ``Online
  graph learning from sequential data,'' in \emph{2018 IEEE Data Science
  Workshop (DSW)}, Lausanne, Switzerland, 2018, pp. 190--194.

\bibitem{dong2019learning}
X.~Dong, D.~Thanou, M.~Rabbat, and P.~Frossard, ``Learning graphs from data: A
  signal representation perspective,'' \emph{IEEE Signal Processing Magazine},
  vol.~36, no.~3, pp. 44--63, 2019.

\bibitem{pasdeloup2017characterization}
B.~Pasdeloup, V.~Gripon, G.~Mercier, D.~Pastor, and M.~G. Rabbat,
  ``Characterization and inference of graph diffusion processes from
  observations of stationary signals,'' \emph{IEEE Transactions on Signal and
  Information Processing over Networks}, vol.~4, no.~3, pp. 481--496, 2017.

\bibitem{thanou2017learning}
D.~Thanou, X.~Dong, D.~Kressner, and P.~Frossard, ``Learning heat diffusion
  graphs,'' \emph{IEEE Transactions on Signal and Information Processing over
  Networks}, vol.~3, no.~3, pp. 484--499, 2017.

\bibitem{chepuri2017learning}
S.~P. Chepuri, S.~Liu, G.~Leus, and A.~O. Hero, ``Learning sparse graphs under
  smoothness prior,'' in \emph{IEEE International Conference on Acoustics,
  Speech and Signal Processing (ICASSP)}, 2017, pp. 6508--6512.

\bibitem{shafipour2017network}
R.~Shafipour, S.~Segarra, A.~G. Marques, and G.~Mateos, ``Network topology
  inference from non-stationary graph signals,'' in \emph{IEEE International
  Conference on Acoustics, Speech and Signal Processing (ICASSP)}, 2017, pp.
  5870--5874.

\bibitem{segarra2016network}
S.~Segarra, A.~G. Marques, G.~Mateos, and A.~Ribeiro, ``Network topology
  identification from spectral templates,'' in \emph{IEEE Statistical Signal
  Processing Workshop (SSP)}, Palma de Mallorca, Spain, 2016, pp. 1--5.

\bibitem{viola2018graph}
I.~Viola, H.~P. Maretic, P.~Frossard, and T.~Ebrahimi, ``A graph learning
  approach for light field image compression,'' in \emph{Applications of
  Digital Image Processing XLI}, vol. 10752.\hskip 1em plus 0.5em minus
  0.4em\relax International Society for Optics and Photonics, 2018, p. 107520E.

\bibitem{sardellitti2016graph}
S.~Sardellitti, S.~Barbarossa, and P.~Di~Lorenzo, ``Graph topology inference
  based on transform learning,'' in \emph{IEEE Global Conference on Signal and
  Information Processing (GlobalSIP)}, Greater Washington, D.C., USA, 2016, pp.
  356--360.

\bibitem{maretic2017graph}
H.~P. Maretic, D.~Thanou, and P.~Frossard, ``Graph learning under sparsity
  priors,'' in \emph{IEEE International Conference on Acoustics, Speech and
  Signal Processing (ICASSP)}, New Orleans, LA, USA, 2017, pp. 6523--6527.

\bibitem{ma2008mining}
H.~Ma, H.~Yang, M.~R. Lyu, and I.~King, ``Mining social networks using heat
  diffusion processes for marketing candidates selection,'' in
  \emph{Proceedings of the 17th ACM conference on Information and knowledge
  management}, 2008, pp. 233--242.

\bibitem{friedman2008sparse}
J.~Friedman, T.~Hastie, and R.~Tibshirani, ``Sparse inverse covariance
  estimation with the graphical lasso,'' \emph{Biostatistics}, vol.~9, no.~3,
  pp. 432--441, 2008.

\bibitem{matta2019graph}
V.~Matta, A.~Santos, and A.~H. Sayed, ``Graph learning with partial
  observations: Role of degree concentration,'' in \emph{IEEE International
  Symposium on Information Theory (ISIT)}, Paris, France, 2019, pp. 1312--1316.

\bibitem{jadbabaie2012non}
A.~Jadbabaie, P.~Molavi, A.~Sandroni, and A.~Tahbaz-Salehi, ``Non-bayesian
  social learning,'' \emph{Games and Economic Behavior}, vol.~76, no.~1, pp.
  210--225, 2012.

\bibitem{nedic2017fast}
A.~Nedi{\'c}, A.~Olshevsky, and C.~A. Uribe, ``Fast convergence rates for
  distributed non-bayesian learning,'' \emph{IEEE Transactions on Automatic
  Control}, vol.~62, no.~11, pp. 5538--5553, 2017.

\bibitem{molavi2017foundations}
P.~Molavi, A.~Tahbaz-Salehi, and A.~Jadbabaie, ``Foundations of non-bayesian
  social learning,'' \emph{Columbia Business School Research Paper}, no. 15-95,
  2017.

\bibitem{molavi2018theory}
------, ``A theory of non-bayesian social learning,'' \emph{Econometrica},
  vol.~86, no.~2, pp. 445--490, 2018.

\bibitem{bordignon2020adaptive}
V.~Bordignon, V.~Matta, and A.~H. Sayed, ``Adaptive social learning,''
  \emph{IEEE Transactions on Information Theory}, vol.~67, no.~9, pp.
  6053--6081, 2021.

\bibitem{bordignon2020social}
------, ``Social learning with partial information sharing,'' in \emph{IEEE
  International Conference on Acoustics, Speech and Signal Processing
  (ICASSP)}, Barcelona, Spain, 2020, pp. 5540--5544.

\bibitem{lalitha2018social}
A.~Lalitha, T.~Javidi, and A.~D. Sarwate, ``Social learning and distributed
  hypothesis testing,'' \emph{IEEE Transactions on Information Theory},
  vol.~64, no.~9, pp. 6161--6179, 2018.

\bibitem{zhao2012learning}
X.~Zhao and A.~H. Sayed, ``Learning over social networks via diffusion
  adaptation,'' in \emph{2012 Conference Record of the Forty Sixth Asilomar
  Conference on Signals, Systems and Computers (ASILOMAR)}.\hskip 1em plus
  0.5em minus 0.4em\relax IEEE, 2012, pp. 709--713.

\bibitem{gale2003bayesian}
D.~Gale and S.~Kariv, ``Bayesian learning in social networks,'' \emph{Games and
  Economic Behavior}, vol.~45, no.~2, pp. 329--346, 2003.

\bibitem{acemoglu2011bayesian}
D.~Acemoglu, M.~A. Dahleh, I.~Lobel, and A.~Ozdaglar, ``Bayesian learning in
  social networks,'' \emph{The Review of Economic Studies}, vol.~78, no.~4, pp.
  1201--1236, 2011.

\bibitem{lalitha2016social}
A.~Lalitha, T.~Javidi, and A.~D. Sarwate, ``Social learning and distributed
  hypothesis testing,'' \emph{IEEE Transactions on Information Theory},
  vol.~64, no.~9, pp. 6161--6179, 2018.

\bibitem{Sayed_2014}
\BIBentryALTinterwordspacing
A.~H. Sayed, ``Adaptation, learning, and optimization over networks,''
  \emph{Foundations and Trends® in Machine Learning}, vol.~7, no. 4-5, pp.
  311--801, 2014. [Online]. Available:
  \url{http://dx.doi.org/10.1561/2200000051}
\BIBentrySTDinterwordspacing

\bibitem{matta2020graphlearning}
V.~Matta, A.~Santos, and A.~H. Sayed, ``Graph learning under partial
  observability,'' \emph{Proceedings of the IEEE}, vol. 108, no.~11, pp.
  2049--2066, 2020.

\end{thebibliography}

\end{document}